\documentclass[10pt, conference, letterpaper]{IEEEtran}

\usepackage{acro}
\usepackage{adjustbox}
\usepackage{algorithmic}
\usepackage{amsfonts}
\usepackage{amsmath}
\usepackage{amssymb}
\usepackage[american]{babel}
\usepackage{booktabs}
\usepackage{chronosys}
\usepackage{epsfig}
\usepackage{graphicx}
\usepackage{hyperref}
\usepackage{cleveref}
\usepackage{listings}

\usepackage{paralist}
\usepackage{pifont}
\usepackage{siunitx}
\usepackage{subfigure}
\usepackage{tabularx}
\usepackage{textcomp}
\usepackage{tikz}
\usepackage{wasysym}
\usepackage{xcolor}
\usepackage{xspace}

\hypersetup{hidelinks=true,breaklinks=true}

\definecolor{base03}{HTML}{002B36}
\definecolor{base02}{HTML}{073642}
\definecolor{base01}{HTML}{586E75}
\definecolor{base00}{HTML}{657B83}
\definecolor{base0}{HTML}{839496}
\definecolor{base1}{HTML}{93A1A1}
\definecolor{base2}{HTML}{EEE8D5}
\definecolor{base3}{HTML}{FDF6E3}
\definecolor{yellow}{HTML}{B58900}
\definecolor{orange}{HTML}{CB4B16}
\definecolor{red}{HTML}{DC322F}
\definecolor{magenta}{HTML}{D33682}
\definecolor{violet}{HTML}{6C71C4}
\definecolor{blue}{HTML}{268BD2}
\definecolor{cyan}{HTML}{2AA198}
\definecolor{green}{HTML}{859900}

\definecolor{darkgreen}{HTML}{237a06}
\definecolor{darkred}{HTML}{bc2e0b}

\newcommand{\eg}{{e.g.,}\xspace}
\newcommand{\ie}{{\it i.e.,}\xspace} 

\newif\iftodo

\newif\ifcomment
\commenttrue

\ifcomment
        \newcounter{MVNumberOfComments}
    \stepcounter{MVNumberOfComments}
    \newcommand{\mvnote}[1]{\textcolor{blue}{\small \bf [MV\#\arabic{MVNumberOfComments}\stepcounter{MVNumberOfComments}: #1]}}
    
      \newcounter{DPNumberOfComments}
    \stepcounter{DPNumberOfComments}
    \newcommand{\dpnote}[1]{\textcolor{magenta}{\small \bf [DP\#\arabic{DPNumberOfComments}\stepcounter{DPNumberOfComments}: #1]}}
\else
    \newcommand\mvnote[1]{}
     \newcommand\dpnote[1]{}
\fi

\newcommand{\etal}[0]{\mbox{\textit{et al.}}\xspace}

\def\BibTeX{{\rm B\kern-.05em{\sc i\kern-.025em b}\kern-.08em
    T\kern-.1667em\lower.7ex\hbox{E}\kern-.125emX}}

\newcommand{\usd}[1]{US\$\,#1\xspace}
\newcommand{\cl}[0]{\emph{c-lightning}\xspace}

\DeclareAcronym{P2P}{
    short = P2P,
    long = peer-to-peer
}

\newcommand{\lnn}[0]{Lightning Network\xspace}

\todotrue
\commenttrue

\begin{document}

\title{An empirical study of availability and reliability properties of the Bitcoin Lightning Network}

\author{\IEEEauthorblockN{Finnegan Waugh}
\IEEEauthorblockA{\textit{University of Sydney}}
\and
\IEEEauthorblockN{Ralph Holz}
\IEEEauthorblockA{\textit{University of Twente} \& \textit{University of Sydney}}
}

\maketitle

\begin{abstract}

    The Bitcoin Lightning network is a mechanism to enable fast and inexpensive
    off-chain Bitcoin transactions using \ac{P2P} channels between nodes that
    can also be composed into a routing path. Although the resulting possible
    channel graphs are well-studied, there is no empirical data on the
    network's reliability in terms of being able to successfully route payments
    at a given moment in time. In this paper we address this gap and
    investigate two forms of availability that are a necessary ingredient to
    achieve such reliability. We first study the Lightning network's ability to
    route payments of various sizes to nearly every participating node, over
    most available channels. We establish an inverse relationship between
    payment volume and success rate and show that at best only about a third of
    destination nodes can be successfully reached. The routing is hampered by a
    number of possible errors, both transient and permanent. We then study the
    availability of nodes in the network longitudinally and determine how
    long-lived they are. Churn in the network is actually low, and a considerable
    number of nodes are hosted on cloud providers. By testing node liveness, we
    find that the propagated network information is relatively often stale,
    however, both for IP addresses and Tor onion addresses. We provide
    recommendations how the Lightning network can be improved, including
    considerations which trade-offs between privacy and decentralization on the
    one hand and reliability on the other hand should at least be reconsidered
    by the community developing the Lightning network.
 
\end{abstract}

\section{Introduction}
\label{sec:introduction}

More than ten years after the creation of the first public blockchain, the
number of transactions that Bitcoin can process is orders of magnitude below
that of classic payment systems operated by banks and credit card providers.
Regulatory issues aside, this is one of the main factors that have held Bitcoin
back from becoming the globally accepted, decentralized cryptocurrency that its
developers meant it to be.  A number of solutions have been proposed to
overcome Bitcoin's performance issues. Among the most interesting ones are
those that focus on bypassing the blockchain for most transactions and use it
as an anchor to keep track only of the result balance of an entire set of
off-chain transactions. Collectively, these approaches are often known as
\textit{Payment Channel Networks (PCNs)}.

The \lnn is such a technology for Bitcoin. It defines \acf{P2P} payment
channels to facilitate smaller transactions between two partnering blockchain
participants. Only the sum of all \ac{P2P} transactions is written out to the
blockchain. By chaining channels together, one can also `route' payments to
participants to which no direct network connection exists. The fees that
channel operators charge are very low compared to Bitcoin's current fees. The
design goals of \lnn emphasize fast payments, with latencies on the order of a
fraction of a second. By keeping most transactions off-chain, the \lnn also
hopes to achieve better privacy for users and a drastic reduction of required
storage space for the blockchain. However, there is a tension between the
design goals of \textit{reliability} and \textit{decentralization}. To be
acceptable to users as a common form of payment, the \lnn must offer a high
degree of reliability in the sense of \textit{payments reaching their
destination}.  An expressive goal of the \lnn is to maintain a high degree of
decentralization~\cite{Poon2016}. As the \lnn is set up as \ac{P2P} network,
where nodes are free to come and leave any time, this means that
\textit{availability}, in different forms, is a prerequisite for the network's
reliable functioning.

In this paper, we focus on two related forms of availability that are a
necessary condition for the latter. The first is \textit{availability of
channels at the time of payment}, \ie the property that (composite) payment
channels must be available and capable of transporting a certain transaction
volume to a payment destination. This property can be directly measured by
running experiments in the network. The second form of availability that we
focus on is \textit{overall availability of nodes in the network}, in the sense
that nodes must not go stale over time: they should remain in the network
for appropriate periods, and they need to respond to queries such as requests
to connect or establish payment channels etc. This second form of availability
is, in fact, a constitutent of the first form; it can also be described as a
function of network composition. However, as we will see, it can be measured
only for a part of the network due to the \lnn's design. This means that it is
important to measure both forms of availability as data for both allows us to
more accurately gauge the \lnn's overall chances of routing payments
successfully. We acknowledge that further factors such as security against
attack, safety of operation such as recovering from errors, etc. are all
further important properties to fully understand the \lnn. They are even harder
to measure at scale, however, and for this paper, we consider them out of
scope.

The \lnn network is now about four years old. Previous work has investigated
the properties of the graph established by payment channels.  However, these
studies are based on \textit{summary snapshots} of channel information that is
propagated in the network over a longer period of time. In this sense, such
studies consider a best-case picture, under the assumption that all channels
and responsible nodes are actually really available. To the best of our
knowledge, there is no existing work that investigates the actual availability
properties of channels at the time a payment is attempted, nor does data on the
overall availability of network nodes exist. In this work, we aim to close this
gap.  Our main contributions are as follows:

\paragraph{Availability of payment routes} We determine the \lnn's ability to
route payments of various, plausible sizes to the intended destination.  We
construct a way to probe the Lightning network for its ability to route
payments of different volumes. Our probing covers almost the entire network.
Our results establish a strong link between success rate and payment volume. We
identify both transient errors (insufficiently funded channels) and intransient
errors (such as offline nodes) as the primary reasons for failing payments. We
find that only small payments have a reasonable chance to be successful and
that only a relatively small number of nodes, and hence payment destinations,
can actually be reached.

\paragraph{Availability of the network substrate} We investigate the directly
observable part of the Lightning network layer, \ie those nodes that publish
their network addresses. While the network is well distributed over many
Autonomous Systems, many nodes are also run in the cloud or hosted by
commercial providers. We test the liveness of nodes and determine network
churn, which is a crucial factor for availability. While the latter is
relatively low, we find that propagated network information is often stale,
despite protocol mechanisms to eliminate it. This contributes to overall poorer
performance.

\paragraph{Recommendations} We conclude with a set of recommendations how the
\lnn can be improved and potentially developed into a viable payment option.
While some changes are mere technical extensions, others require to rethink the
degree of privacy desired for the network and the level of centralization that
one wants to accept.

The remainder of this paper is organized as follows. In the next section, we
provide the relevant background to understand how the \lnn works and is set up.
We follow this up with a discussion of related work in \Cref{sec:relwork}. In
\Cref{sec:methodology}, we describe our choice of measurement methods with
respect to determining availability and reliability. We present our results in
\Cref{sec:results} and give key take-ways. We discuss potential improvements,
including a rethinking of privacy and centralization aspects, in
\Cref{sec:discussion}.

\section{Background}
\label{sec:background}

The Lightning network is a representative in the class of Payment Channel
Networks (PCNs), which are one possible answer to the scalability issues
inherent to many blockchain designs.  Bano \etal provide a good introduction to
blockchain and the particular problem space of accelerating payments
in~\cite{Bano2017}.  Several different approaches to PCNs are known; Gudgeon
\etal give an overview in~\cite{Gudgeon2020}.

\paragraph{Payment channel networks} In PCNs, a large number of transactions
between a limited number of nodes can be carried out `off-chain', without
sending the transactions to the blockchain.  Conventionally, the blockchain
consensus protocol is responsible for deciding which transactions are
considered correct and committed to the blockchain; herein lies an important
bottleneck. In a PCN, the consensus protocol is typically only used to anchor
the faithful creation and deletion of \emph{payment channels} between nodes. At
the time of writing, only Bitcoin's \lnn is operated on a main blockchain by
default. In the following, we limit ourselves to the \lnn.

\paragraph{Difference to blockchain network} Since PCNs operate on top of a
blockchain, they are sometimes called Layer 2 solutions, built on top of the
blockchain consensus protocol (Layer 1), which itself runs over \mbox{Layer 0},
which is the protocol that organizes the blockchain \ac{P2P} network. We do not
use the layer terminology here to avoid confusion: two distinct protocol layers
can also be identified for the \lnn itself.  One is responsible for the
creation and management of a \ac{P2P} network of participating Lightning nodes,
and one is used to manage payment channels and forward payments. Consequently,
the substrate of the \lnn is a \ac{P2P} network that is distinct from the
Bitcoin network that maintains the blockchain: every \lnn node runs a Bitcoin
client as well, but the inverse is not true. The wallet used for the Bitcoin
client is also different from that for the Lightning client. The network
organization happens separately, but in a similar way, using a discovery
protocol based on a gossiping mechanism. Nodes in the \lnn have identifiers
that are used in the Lightning protocol. These are propagated between
participating nodes. However, the \textit{network} addresses, \ie IP and port
or alternatively Tor onion addresses, are \textit{not} necessarily gossiped.  A
client participating in the \lnn learns only the network addresses of nodes
that choose to reveal them.  In the popular \emph{c-lightning} implementation,
for instance, operators must manually enable this. This is ostensibly done for
privacy reasons. However, it means that many nodes that are potential
destinations are only known via their identifier, and they can only be reached
via the node(s) that they choose to connect to themselves. All other nodes in
the network must hence find a \textit{route} via the latter nodes to send them
a payment.

\paragraph{Payment channels} A node that is connected to another Lightning node
is expected to open a payment channel. In essence, this is an agreement between
the two nodes to commit a certain amount of Bitcoin to the operation of this
channel. This initial commitment (the funding transaction) is anchored in the
Bitcoin blockchain with a special transaction, where the funds are stored in a
so-called multisignature address that is controlled by both parties. Each party
has a balance in the channel, and both balances together make up the
\emph{channel capacity}. The key idea is now that both parties can send each
other any number of transactions (so-called commitment transactions) and update
their respective view of the balances, with newer commitment transactions
invalidating previous ones. As long as both parties behave honestly, there is
no need to involve the blockchain consensus protocol. This only needs to happen
when the channel is closed again: the peers are reimbursed and the final
balances are written out to the blockchain. A channel can be closed
unilaterally or bilaterally.  Commitment transactions are still normal Bitcoin
transactions; the obvious risk to address is the misbehavior of a peer
attempting to steal funds. Lightning does this with a relatively complex
mechanism that ensures that only the most recent commitment transaction is ever
broadcast to the blockchain \textit{and} accepted.  Each peer holds
cryptographic proof that can be used against the other side in case this
condition is violated, \ie one party sent an older commitment transaction to
the blockchain. In such a case, the funds of the misbehaving side are awarded
to the victim of the attempted fraud. The exact mechanics are described in a
whitepaper~\cite{Poon2016}; implementation details are given in an RFC-style
document~\cite{LightningRFC}.

\paragraph{Sending payments} Payments in Lightning are fundamentally different
from Bitcoin transactions. A receiver must create a so-called \emph{invoice}
and make it available to the sender. With the invoice, the sender can try to
make the payment. Channels may be chained to allow sending payments to any
destination to which a valid route (chain of channels) can be determined. This
is achieved with so-called Hashed Timelock Contracts. The idea here is that the
sender determines a path over a number of channels (source routing).  Each node
on this path is only aware of predecessor (incoming payment) and successor
(outgoing payment); this is the same principle as in the Tor network (onion
routing). To set up the entire routed payment, the node that is the final
destination of a payment generates a secret, random value and sends it as a
hash value to the source of the payment as part of the invoice. A chain of
payments along the channels is created based on this hash value. Once the
destination redeems the final transaction, the preimage of the hash is revealed
in such a way that, in reverse, every node on the path can also redeem the
funds they used to make the payment possible. The invoice mechanism provides a
very useful way for us to probe the network: by generating invalid invoices, we
can determine the validity and capacity of routes without a need to make actual
payments.

\paragraph{Routing} To allow nodes to identify paths through the network, the
\lnn broadcasts every known channel between nodes. This only needs to involve
the channel identifiers and the identifiers (public keys) of source and
destination nodes, not their network addresses. The balances of channels are
not broadcast for privacy reasons.  Consequently, it is entirely possible that
a routed payment fails because one of the channels in the chain does not have
the necessary capacity or the sender is currently insufficiently funded as the
channel has already executed too many outgoing transactions. This is signalled
back to the source with an error code once the payment is tried. We make use of
this behavior when probing the network: it allows us to test which transaction
volumes are supported by the network and how many nodes can be reached in
relation to the transaction volume.

\section{Related work}
\label{sec:relwork}

While the Bitcoin network and a few other blockchain networks have been the
subject of measurement studies before (\eg ~\cite{Decker2013, Gencer2018,
Kim2018}), PCNs have received less attention.  Several websites exist that
visualize the information that can be obtained from running a Lightning node
and offer so-called `public snapshots' (\eg \url{1ml.com},
\url{lndexplorer.com}). Such data is used in a number of publications,
\eg~\cite{Martinazzi2020,Ersoy2020,Beres2019,Seres2019}.  A drawback of this
kind of data source is that it is unclear whether outdated information
(inactive nodes) has been removed; it is also unknown from which or how many
vantage point(s), and at which time, the data was raised.

To the best of our knowledge, there is no previous academic publication that
analyzes the composition of the Lightning \ac{P2P} network or identifies the
\lnn's efficacy in terms of routing. The work that is probably closest to ours
is published by Decker in the Blockstream company blog~\cite{Blockstream2019}.
The author describes a reachability experiment that also uses the \emph{probe}
module of the \emph{c-lightning} implementation. However, the probing is
limited to one-hop reachability tests. The author identifies just 829 nodes
with publicly announced network addresses. At the time of measurement (November
2019), 27.4k channels were known, with a total capacity of just 827 BTC.
Disregarding nodes without active channels, Decker identifies 65\% of nodes as
reachable, \ie responsive to the probe and a working destination for payments.
He also briefly investigates transaction latency and stuck payments.  Unlike
our work, there is no systematic attempt to explore the capability of the
network to transport payments of various volumes nor reachability over routes
consisting of several channels.

Most related work studies only the channel graph that is propagated in the
network, but not actual availability by measurement. The graph's robustness
properties have been investigated several times, including its small-world and
scale-free properties. Martinazzi and Flori study the graph for the first year
of the \lnn's existence~\cite{Martinazzi2020}, relying on public snapshots and
determining graph properties, especially scale-freeness.  Seres
\etal~\cite{Seres2019} use public snapshots from January 2019.  At the time,
the channel graph consisted of 16.6k channels; the \lnn had a capacity of 540
BTC.  Naturally emerging scale-free networks are typically very robust to
random failure but not to targeted attack. The authors confirm this for
Lightning as well by simulating the removal of nodes with many channels. An
important conclusion is that the network needs dedicated protection against
Denial-of-Service attacks where the attackers go after the most important
nodes. Similar results are also presented in~\cite{Guo2019,Lee2019,Lin2020},
with the latter concluding that the Lightning network exhibits a core-periphery
structure.

Rohrer \etal~\cite{Rohrer2019} derive the channel graph from running two nodes
in the network themselves and dumping the identified channels. They also
establish the small-world property and the scale-freeness of the graph but
bolster their result with robust statistical tests, which is an absolute
requirement to identify scale-free networks\footnote{It is a premature conclusion to call a
graph scale-free based only on visual inspection of a double-log plot as many
other authors did; see~\cite{Broido2019} for details.}. The authors provide a
count of vertices in the channel graph, which rose from 1.5k to 2.4k between
November 2018 and February 2019.  They then design an attack to make channels
temporarily unavailable by creating HTLCs such that nodes on a route must wait
for the time locks to expire.

Herrera-Joancomarti \etal~\cite{HerreraJoancomarti2019} design an attack to
determine the balances of Lightning channels, which are meant to be
confidential between the endpoints to increase privacy. They show a
proof-of-concept on the Bitcoin testnet that involves the repeated probing of a
channel with the help of invalid payments.  While the design of the probing is
similar to our approach, we are not concerned with attacks in this paper.
In~\cite{Sola2020}, the same authors build on the above attack to temporarily
lock the channel balance of targeted, central nodes, making them unavailable
for the rest of the network.

Some work analyzes the economics of operating a Lightning node. Ersoy
\etal~\cite{Ersoy2020} design an algorithm to maximize profit from operating a
Lightning channel. They evaluate their algorithm using empirical data
downloaded from \url{1ml.com}. In \cite{Beres2019}, the authors investigate
whether running a \lnn node is economically rational. They simulate
transactions on the \lnn based on graph snapshots and data on transaction fees
from~\cite{Rohrer2019}. Based on the current channel graph and some (strong)
assumptions of user behavior, they answer the question in the negative.

Finally, several authors devise solutions to remove potential bottlenecks in
payment routing, \eg \cite{Roos2018, Miller2019}. These are not necessarily
specific to the Lightning network.

\section{Methodology}
\label{sec:methodology}

We explain our methods to test the availability of payments routes and the
overall node availability in the network.

\subsection{Availability of channels payment routes}
\label{sec:sub:efficacy}

The first part of our methodology addresses the first form of availability we
described in \Cref{sec:introduction}. The goal is to measure systematically how
many payment channels and payment routes support a payment of a given volume
and how many destination nodes can be reached by routing payments through the
network, over channels of various lengths.

We test channels by running a Lightning node in the cloud, specifically a
DigitalOcean location in San Francisco. We refer to this node as Node A. The
node has both IPv4 and IPv6 connectivity.We fund our own node with the
equivalent of about US\$\,120 and open initial channels to well-connected
nodes. We choose the two nodes that have the highest number of public open
channels and do not reject our request to create a channel with sufficient
funding. One node accepts our funding of the Bitcoin equivalent of \usd{100};
the other only accepts a funding of \usd{20}.  At the start of our probing, one
node is the source point of 867 channels; the other of 840.

\emph{c-lightning} provides the ability to add user-defined plugins to extend
its functionality. One such plugin is the \textit{probe} module, which allows a
user to test a payment to another node. It achieves this by sending a payment
with an invalid (random) payment hash. When the payment arrives at its
destination, the recipient is unable to redeem it and returns an
\textit{Unknown Payment Details} error. This error indicates that the
reachability experiment was actually successful. If the payment cannot be
routed through the specified route, then a different error will be returned
(see below). An advantage of this kind of probing is that it allows to test
payments without actually spending Bitcoin.

We modify the probe module to support testing the availability of channels for
payments of a certain volume. We do this by breadth-first search. Using the
method described above, the modified plugin tests payments through as many
channels as it can reach.  It first tries to route a payment through the
channels it has open with the initial peers. If these are successful, it
attempts to route the payment through each of the channels that each of the
respective peers has open, except for those channels that have already been
tested. It continues in this breadth-first manner until no more channels can be
reached. We cannot probe every single channel in existence: some are
unreachable from our initial nodes. To test them, we would have to iteratively
establish channels to (potentially many) other nodes. As there is no guarantee
they would accept payments of the volumes we choose, we regard this alternative
methodology as having diminishing returns. Furthermore, we note that payments
may also fail due to transient reasons (see \Cref{sec:sub:channels}).

We begin our experiment on 2019-11-03 and end it on \mbox{2019-11-25}. For each
attempted payment (reflecting a composition of channels), we make at most two
payment attempts and store whether one of the two attempts was `successful' or
we could not recover from an error.  Since each channel direction has a
different balance and potentially different fees for forwarding, we probe each
channel direction. As transaction volumes, we choose the Bitcoin equivalent for
\usd{0.01}, \usd{1}, \usd{5}, \usd{10},  \usd{50} and \usd{100} at the time
of the experiment's start.

\subsection{Availability of nodes in the \ac{P2P} network}
\label{sec:sub:measuringp2p}

The second set of measurements addresses the second form of availability we
discussed in \Cref{sec:introduction}. The goal is to establish how many nodes
whose network addresses are propagated through the network actually reply to
attempts to contact them. This also allows us to compute churn in the network.
As explained in \cref{sec:background}, measurements of the Lightning network
layer are limited to nodes that consent to the propagation of their network
addresses by other peers.

\subsubsection{Input data}

We have two input data sources, which we use for different purposes. First,
we have access to data raised by developers of the \cl client. The
\emph{listnodes} RPC call provides a mapping between the \emph{node identifier
(node ID)} and known network address.  These can be any combination of IPv4,
IPv6, and Tor onion addresses. The data set is split into days. For each day,
it lists the node IDs and last known addresses on the day or prior. The data
set covers the period 2019-09-15 through 2019-11-30. In addition to node information,
we also obtain the known channels (with source and destination nodes) of each
day.

According to~\cite{Blockstream2019}, stale nodes, \ie nodes with no active
channels, should be purged with the client's default settings. This view of the
network would be in contrast to public visualization sites such as
\emph{1ml.com}, which do not employ this purging and consequently overestimate
the network size. The specification of the \lnn protocol also mentions pruning
stale nodes and channels explicitly. We would hence expect most nodes to be
responsive. In order to test the liveness of nodes ourselves, we install a
further Lightning node in the network of the University of Sydney, which we
refer to as Node B.  The university network has only IPv4 available. We fund a
channel with 0.001 BTC, \ie about \usd{10} at the time of the experiment. We
run the node from 2020-02-18 onwards. As the node has only one channel, we note
that this may mean the node receives fewer announcements via the dicovery
gossip protocol than nodes with many active channels (and hence connected
peers). We return to this in \Cref{sec:sub:network}. We also use the
\emph{listnodes} RPC call of \emph{c-lightning} to dump node information.

We map the IP addresses of Lightning nodes to Autonomous Systems (ASes) using
\emph{pyasn} in conjunction with Routeviews routing information at the
corresponding time. We use Team Cymru's WHOIS service
(\url{v4.whois.cymru.com}) to map the AS number to the operator name and
registration with a Regional Internet Registry (RIR). This gives us the name of
the network hosting a node.

\subsubsection{Testing liveness}

For nodes learned with our own Lightning Node B, we use two methods to check
their liveness and reachability. For \emph{IPv4 nodes}, we use the extremely
fast Internet scanner \emph{zmap}~\cite{durumeric2013zmap} to test whether they
respond on the allocated port. This scan takes place from the University of
Sydney and takes only two minutes.  Almost all nodes use the Lightning default
port (9735); we hence limit our investigations to this port. 

A significant number of nodes publish a \emph{.onion} network address, which
refers to a Tor hidden service. Such nodes are only reachable via rendezvous
points using the Tor anonymization network. We deploy a Tor daemon on our
server and connect via the \cl client to test their liveness.

\subsubsection{Limitations} Only a subset of nodes publishes their network
addresses.  This may introduce a bias: it is conceivable, for example, that
operators running the Lightning software on a public server (\eg a VPS) are
more likely to publish their network address than private users in a home. This
is particularly true if they aim at making a profit by collecting fees from
their channels. A similar argument holds for Tor hidden services, which should
ideally also be longer-lived. We keep this limitation in mind when discussing
our results.

\subsection{Ethical considerations and reproducibility}

Although it is deployed on Bitcoin's \emph{mainnet}, the \lnn network is still
in a testing stage. Developers monitor the network and occasionally probe it
for metrics like reachability~\cite{Blockstream2019}. This monitoring is one of
our data sources. Internet scans and interactions with cryptocurrency networks,
which constitute another part of our methodology, have been cleared by the
Human Ethics Committee of the University of Sydney. We employ several guiding
principles to minimize the impact of our measurement on the \lnn.  We refrain
from attempts to include every single channel in our breadth-first search as
this would imply many additional channels would have to be opened.  We also
limit ourselves to just two probes for payments.

We support open science and will make both our source code and our data
available publicly available.

\section{Results}
\label{sec:results}

We first present our results on availability of payment routing and then the
results on overall availability of nodes in the network.

\subsection{Availability of channels}
\label{sec:sub:channels}

According to our data source from the \cl developers, which collected daily
snapshots, between 31.4k and 31.8k channels existed between 2020-11-03 and
2020-11-25 and were not stale. Although we do not attempt a full combinatorial
testing of all possible routes (see \Cref{sec:methodology}), our own probes
from our initial, well-connected hop cover a significant amount of the
propagated channels: 30.7k are reached by the probes. This corresponds to 49.2k
channel directions. Note that we do not reach the theoretically possible 61.4k
directions: many nodes have only a single channel open, and if a probe is not
able to traverse the channel to a given endpoint, then it is not possible to
test the reverse direction, either. While the overall number is high and seems
to imply that reachability in the \lnn is very good in principle, the picture
changes dramatically when looking at the success rate of payments.

\subsubsection{Success of payments}

According to our data from the \cl developers, between 4826 and 4950 nodes are
observed during our experiment period. In our experiments, we attempt payments
to a total of 4626 unique nodes, \ie we cover between 93.5\%-95.9\% of existing
nodes. However, the success rate of attempted payments varies between 15.32\%
and 72.09\%, depending on the payment volume. The number of nodes that we reach
with a payment is also considerably lower than the number of nodes we attempt
to send a payment to: just 2055. This number is the \textit{union} of all
unique nodes reached over all probes of various payment volumes
(\Cref{tab:successratebyvolume}). We can never reach more than roughly at third
of all nodes, independent of payment volume. Note that the relatively long
duration of our experiment implies that some nodes that were reachable during
tests with one payment volume may be unreachable again for the next, or vice
versa.

We observe a strong inverse relationship of successful payments to the volume
used, which we summarize in \Cref{tab:successratebyvolume}. While by far the
most very small payments of \usd{0.01} are successful overall (72\%), we still
reach only 35\% of nodes overall. The numbers for successful payments drop
immediately when we increase the volume to \usd{1}, however the number of
reached nodes stays almost the same. Evidently, a sufficient number of payment
channels are alive and functioning to enable this volume. While 35\% of nodes
seems to be an upper number in terms of reachability for practically useful
payments, we note that some of the payment routes are longer than others,
meaning that such routes imply a higher overall chance of failure. The numbers
deteriorate fast with higher payment volumes.  Payments of \usd{50} and above
do not reach any significant number of nodes any more.  It is unclear from our
observations why more nodes can be reached for \usd{10} than for \usd{5};
however, we emphasize the high rate of transient errors in channels, which we
explore below.

\begin{figure}[tb]
    \centering
    \includegraphics[width=0.40\textwidth]{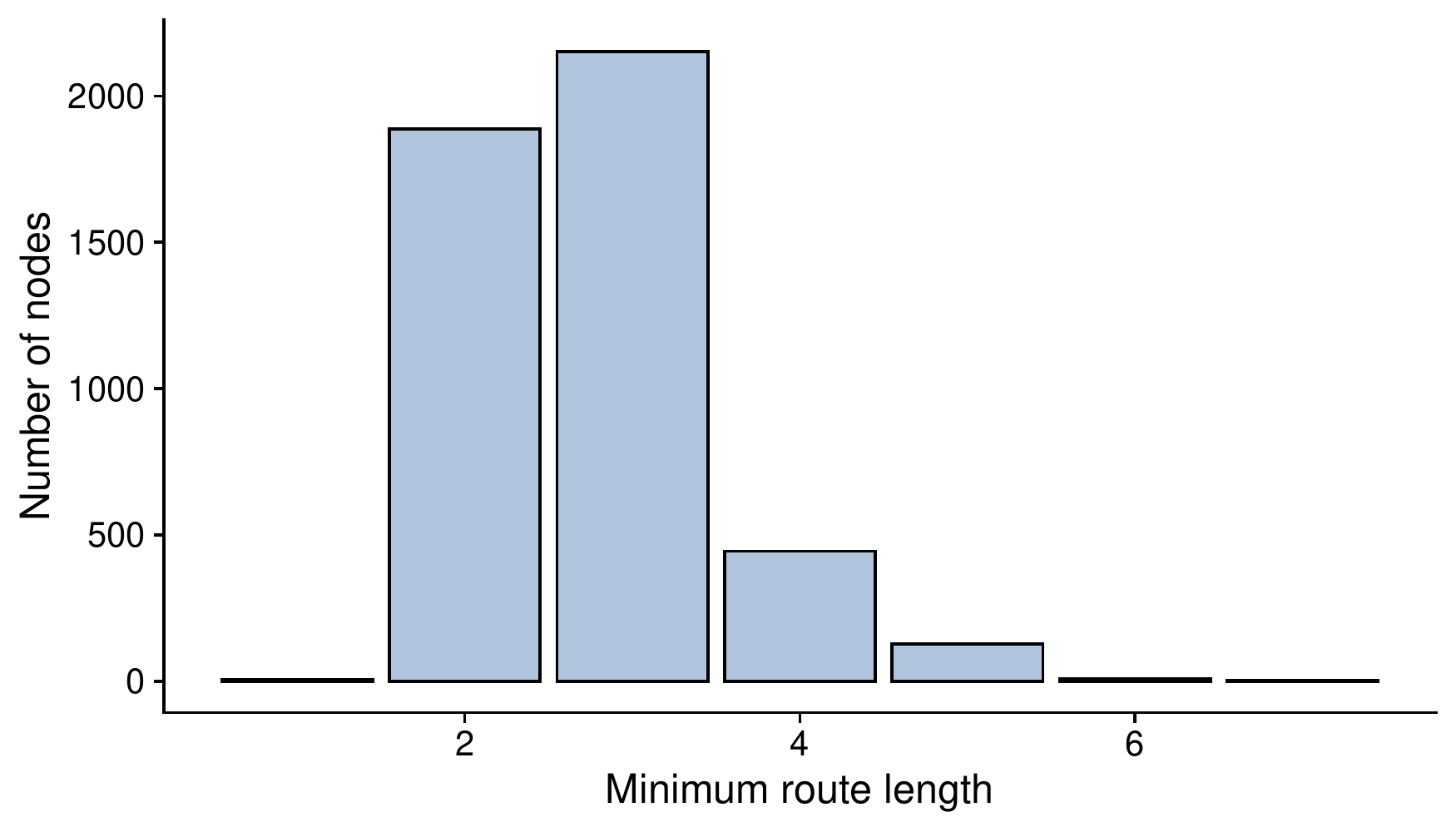}
    \caption{Minimum number of hops to reach nodes.}
    \label{fig:route-length}
\end{figure}

We also compute how many hops are needed to reach the destination.
\Cref{fig:route-length} shows this. Even with just two outgoing channels, it is
possible to reach the vast majority of nodes via two or three hops. Note that
the principle of onion routing requires two hops from the source, \ie one
`intermediate' and one `exit' node before the final hop (the destination).

\begin{table}
    \centering
    \caption{Success rate for payments and \textit{unique} nodes reached.}
    \label{tab:successratebyvolume}
    \begin{tabular}{lrrr}
        \toprule
        Amount (\usd{})     & \% payments   & \% nodes & \# nodes \\
        \midrule
        0.01        & 71.92         & 35.24 & 1630 \\
        1           & 57.82         & 35.19 & 1628 \\
        5           & 44.06         & 25.66 & 1187 \\
        10          & 44.15         & 33.72 & 1560 \\
        50          & 30.93         & 9.22 & 459 \\
        100         & 17.30         & 4.82 & 223\\
        \bottomrule
    \end{tabular}
\end{table}

\subsubsection{Reasons for failing payments}

\Cref{tab:errorbreakdown} summarizes the errors we determined across all
payments. The most common error was a temporary channel failure.  This error is
reported when a channel direction has temporarily insufficient funds available.
This can be a result of previous transactions through the channel that have
been locked. The error indicates the same payment may succeed through the
channel at a later time. The next most common error was \emph{Unknown Next
Peer}. This indicates a hop along the intended payment route did not have a
connection to the next node. Less than 1\% of payments returned a channel
disabled error.  Other errors occurred very rarely; we attribute them to the
client having slightly out of date information about some parts of the network.
Payment timeouts are particularly rare. These errors indicate that the payment
has taken longer than four minutes to return. When these errors occur they lock
the funds in place untl the payment returns, and in some situations channels
have to be closed to recover the funds. This can be a costly (due to Bitcoin
transaction fees) and lengthy process.

\begin{table}
    \centering
    \caption{Breakdown of errors.}
    \label{tab:errorbreakdown}
    \begin{tabular}{lrr}
        \toprule
        Result & Count & \% \\
        \midrule
        Temporary Channel Failure & 47341 & 31.58 \\
        Unknown Peer & 23839 & 15.90 \\
        Channel Disabled & 1203 & 0.80 \\
        Payment Timeout & 336 & 0.22 \\
        Insufficient Fee & 192 & 0.13 \\
        Expiry Too Soon & 142 & 0.09 \\
        Other & 267 & 0.18 \\
        \bottomrule
    \end{tabular}
\end{table}

\begin{figure}[tb]
    \centering
    \includegraphics[width=0.50\textwidth]{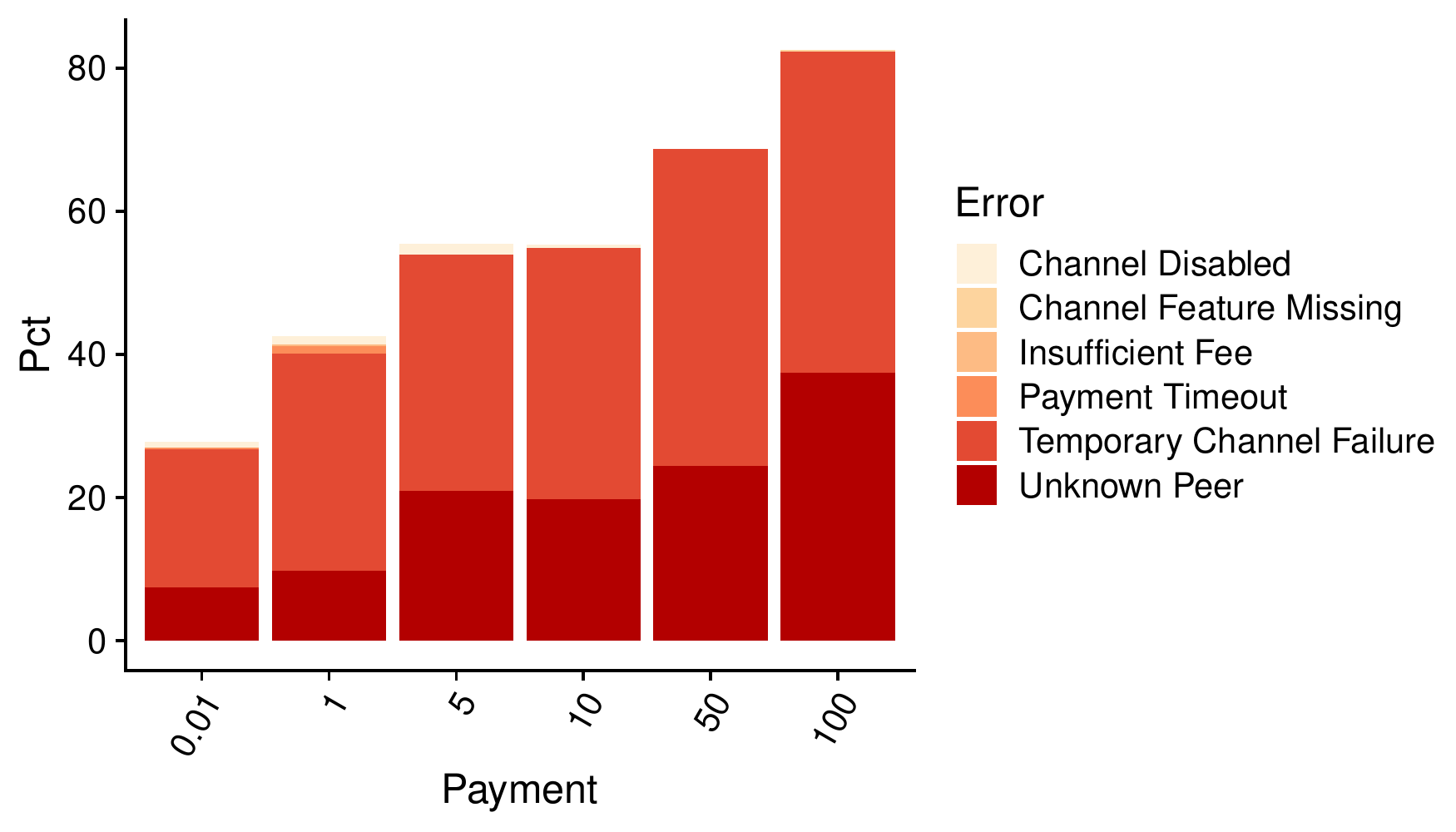}
    \caption{Errors during payments broken down by payment volume (\usd{}).}
    \label{fig:errors-by-volume}
\end{figure}

\Cref{fig:errors-by-volume} breaks the errors down by payment volume, grouping
rare errors together. While the percentages of errors vary by volume, the
relative proportions are very similar between volumes.  We find that probed
channels generally either do not return the \emph{Unknown Peer} error at all,
or they return it consistently. Of the more than 30k channels we probe, 6550
return the error on every probe. This suggests that over the probing period of
a couple of weeks, the nodes at either ends of these channels were either
infrequently or never connected. Similarly, we also find that there are 1780
nodes over which we cannot route a single payment.

These numbers seem so high that we verify them with an alternative client after
our main measurements. We temporarily connect a second client to the network
and retry payments to the endpoints of channels where we received the
\emph{Unknown Peer}. The second client is successful in routing payments to
about 20\% of these nodes. While this seems a slight improvement, it still
validates the overall frequency of this error.

We note that there is a risk for nodes that remain disconnected over longer
periods of time while keeping channels open. The respective endpoints of the
channels may unilaterally try to close the channel and publish a signed
transaction with an old channel balance. If this remains undetected for a
longer period of time (for example, 24 hours), then the funds may be stolen
without recourse. The concept of `watchtowers' has been proposed to warn
potentially affected
nodes\footnote{\url{https://hedgetrade.com/what-are-lightning-networks-watchtowers/}}.

\textbf{Key take-aways.} Although many channels are reachable in principle, the
\lnn is actually not capable of routing larger payments. Even modest payments
of just US\$\,10 succeed only in less than half of all attempts; for larger
payments around US\$\,50, the success rate drops to below 20\%. No experiment
run was ever able to reach more than a third of the destination nodes. At
present, the \lnn is not capable of supporting the kinds of payments that would
be needed to support activities such as shopping for everyday goods. This is
due to insufficient funding of channels: even though many channels exist, they
are often not able to forward a payment due to \textit{congestion}: at the time
we attempt a payment, it exceeds the channel's currently available balance. A
further, important reason is that the node corresponding to a channel's
endpoint is unavailable, often for long periods of time. This suggests that
channel management could be improved.

\subsection{Overall node availability on the network}
\label{sec:sub:network}

More nodes in the network imply more available channels and a higher overall
availability. We hence determine the network's growth over time. Node churn is
an important factor, however, as nodes should ideally stay well-connected in
the network to sustain the overall reachability. We hence also measure how
long-lived nodes are. Network location is an important factor related to this:
nodes that run on publicly reachable servers can be expected to contribute more
to overall reachability due to their uptime.

\subsubsection{Network growth}

\begin{figure}[tb]
    \centering
    \includegraphics[width=0.50\textwidth]{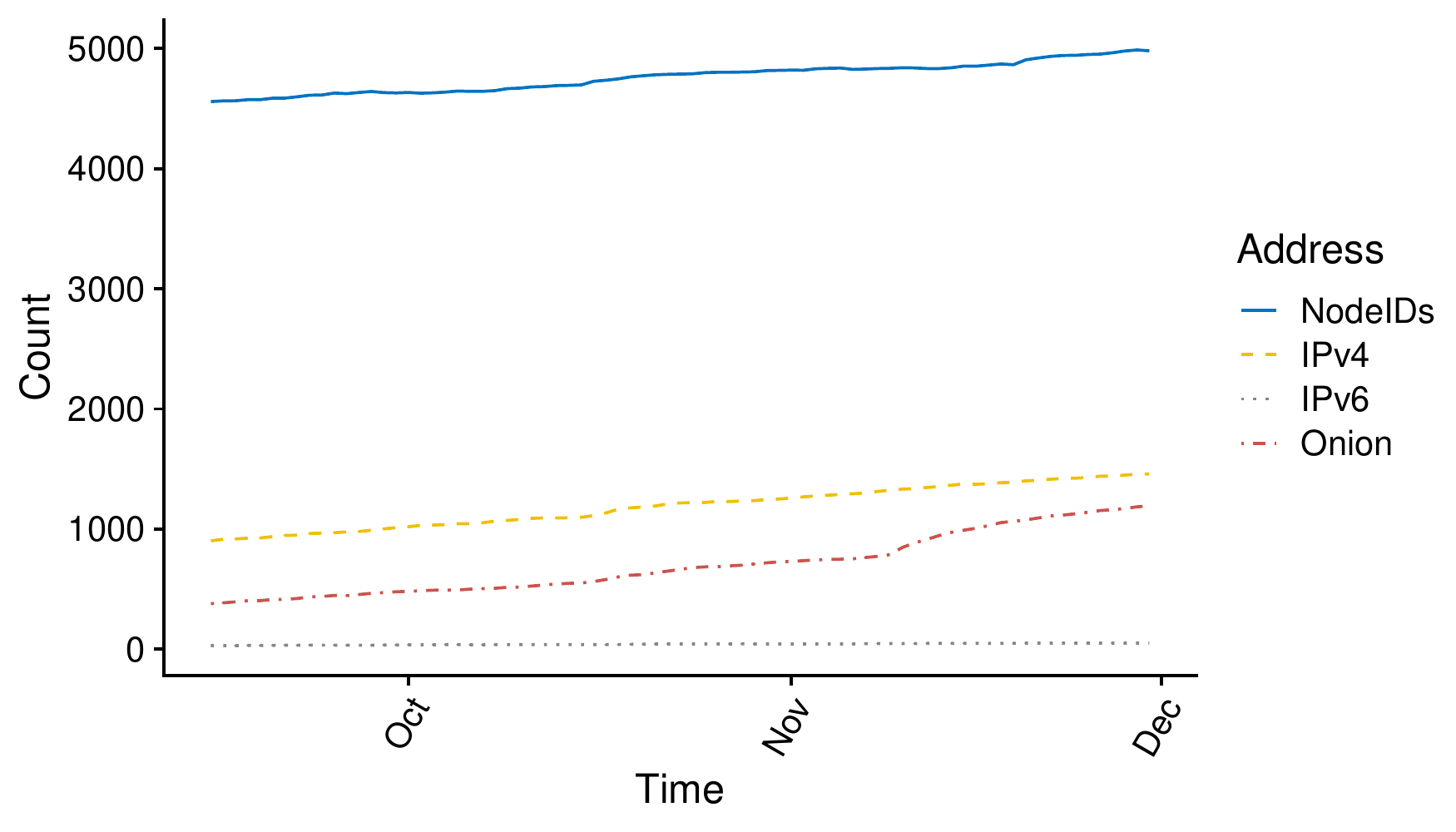}
    \caption{Development of the network layer: node IDs and network addresses.}
    \label{fig:blockstream-development}
\end{figure}

\Cref{fig:blockstream-development} shows the development of the \lnn according
to the data from the \mbox{\cl} developers. We identify a slow but stable
growth between \mbox{2019-09-15} and \mbox{2019-11-30}: from 4558 Node IDs to
4981, \ie nearly 10\% in 2.5 months. This corresponds to significantly fewer
network addresses: just over 1000 IPv4 addresses are publicly known, and nearly
as many onion addresses. This means many nodes do keep the privacy feature of
not propagating their network address enabled.

Interestingly, IPv6 addresses are rare---we never find more than about 50.
Using a second protocol can potentially improve overall availability. A number
of nodes publish an IP address \textit{and} an onion address. For example, 79
nodes choose to do this on 2019-09-15.  This numbers grows to 182 by
2019-11-30. However, doing this breaks the anonymity of the respective NodeID
as a receiver of payments. Either operators are not aware of this, or
indifferent to it, or their primary goal is not their own anonymity but merely
to help making forwarded payments as anonymous as possible.

We also determine the network addresses from the point of view of our Node B as
of 2020-02-18 12:00 GMT, \ie 2.5 months later. At the indicated time, our node
is aware of 4938 unique nodes (identified by their node ID). If we assume the
\lnn maintained its growth after 2019-11-30, and assuming the growth rate
remained about the same, this would imply our node is aware of about 90\% of
nodes in the network. For the 4938 nodes that Node B is aware of, we identify
2296 unique IPv4 addresses (a further 45 are in IANA-reserved prefixes, \ie
invalid).

Our data from 2019 shows a relatively high number of onion addresses in the
network until 2019-11-30. Our Node B identifies unique 1885 onion addresses on
\mbox{2020-02-19}, which indicates a definite growth since the end of our first
observation period on 2019-11-30.

Our node set of 2020-02-19 11:30 GMT, which we obtain via the \emph{listnodes}
RPC call, contains 1882 onion addresses corresponding to 1843 nodes.  410
addresses use the older v2 addresses. These addresses use SHA1 and RSA-1024
cryptography, which offer less security against impersonation attacks and are
slowly being phased out.  Hidden services of v2 are also known to leak more
information to Tor's hidden directory services, an undesirable property.
However, overall the choice of v2 addresses is not too concerning at this time,
and the fact that the majority uses the newer v3 services is in fact
encouraging.

\subsubsection{Liveness and churn}

We first use our data set from 2019 to check how long-lived nodes are.
Comparing the node IDs from 2019-09-15 to the ones from 2019-10-15, we find
that many remain in the network. Of 4558 nodes on 2019-09-15, 4038 are still
there one month later. A further month on, this number drops to 3197, however.

Using our Node B, we estimate liveness and churn over the node's observation
period.  Note that we do this for network addresses, not the public key-based
identifiers. Of the 2251 public IP addresses where the Lightning software runs
on port 9735, 73 are in IP ranges that are on our blacklist of networks that do
not wish to be scanned. We choose the remaining 2178 addresses to scan TCP port
9735 and test their liveness. Only 875 nodes, corresponding to 859 IP
addresses, respond on the port. This is a surprise as we use a version of \cl
that is meant to prune stale nodes; hence the client should have a much more
accurate picture of the network.

We compute the IPv4 churn. We use the node set of Node B at the same time
(2020-02-18 12:00 GMT) as the reference set of live and unreachable nodes. We
then compare this to the node sets at the following times: 10, 30, and 60
minutes later; 2, 4, 8, 12, 16, 20, and 24 hours later; and 36, 48, 60, and 72
hours later.

\begin{figure}[tb]
    \centering
    \includegraphics[width=0.50\textwidth]{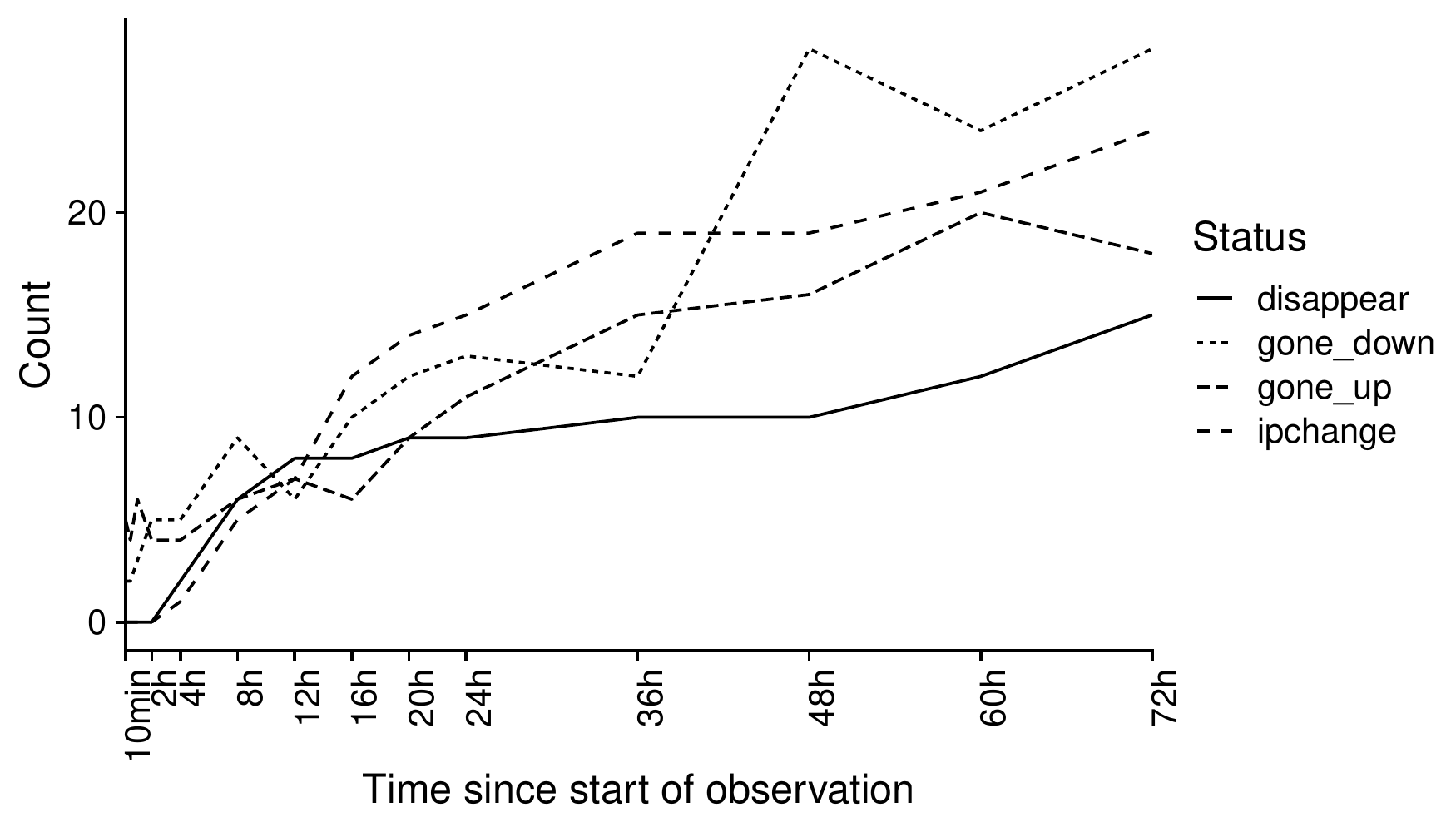}
    \caption{Network churn within three days.}
    \label{fig:churn}
\end{figure}

We find that there is very little change in the network in terms of nodes
remaining reachable or unreachable. Although two nodes become unreachable in
the first ten minutes, 837 nodes are still reachable under the same IP address
and port even after 72 hours (more than 95\%). \Cref{fig:churn} summarizes
changes that occur over time. In the first 8 hours, there is very little
activity: a handful of nodes (by node ID) disappear from the node set or become
unreachable, but similar numbers also become reachable again. Larger changes
become apparent only later. There is a jump in nodes that become unreachable
after 12 hours, although a similar jump also occurs for nodes that become
reachable. Overall, the network seems quite stable, and the churn is not a
strong explanation for the observed stale addresses.

We choose the onion nodes for a manual liveness test as well. We enable use of
the Tor daemon for our \cl client and attempt to connect to each onion address
in turn.  This is a lengthy process (13 hours). We successfully connect to 1297
of the addresses and fail in 546 cases. This indicates that onion addresses
also suffer from the problem of staleness.

\begin{table*}
    \centering
    \caption{Autonomous Systems containing Lightning nodes with public IPv4 addresses, with Top 5.}
    \label{tab:ases}
    \begin{tabular}{lrrrrrrr}
        \toprule
        Time    & IPs & Spread & 1st (\# nodes) &  2nd (\# nodes) & 3rd (\# nodes)     & 4th (\# nodes) & 5th (\# nodes) \\
        \midrule
        2019-09 & 901 & 258 ASNs & Google (116)   &  Comcast (50)   & DigitalOcean (45)  & Cogent (37)    & OVH (21) \\
        2019-10 & 1.1k & 305 ASNs & Google (121)   &  DigitalOcean (57) & Comcast (56) & Cogent (49) & Amazon (28) \\
        2019-11 & 1.3k & 337 ASNs & Google (123)   &  DigitalOcean (70) & Cogent (63) & Comcast (62) & Amazon (35) \\
        2020-02 & 2.3k & 471 ASNs & Cogent (141)   &  Comcast (133) & DigitalOcean (116) & Amazon (78) & Google (78) \\
        2020-02 (live) & 858 & 249 ASNs & Cogent (114)   &  DigitalOcean (59) & Amazon (38) & OVH (32) & Contabo (29) \\
        \bottomrule
    \end{tabular}
\end{table*}

\subsubsection{Network locations}

We map the public IPv4 addresses to their corresponding ASes. We choose the
18th of \mbox{2019-09}, \mbox{2019-10}, \mbox{2019-11}, and \mbox{2020-02} for this analysis.
\Cref{tab:ases} presents the results. We first observe that the Lightning IP
addresses are spread over a wide range of Autonomous Systems Numbers, and this
number is increasing. In 2019-09, about 900 Lightning IPs are distributed over
more than 250 ASNs (although we note that large corporations sometimes
hold several ASNs). The top 5 AS owners between September and November 2019 are
largely the same. We find hosting/cloud providers such as Google, DigitalOcean,
Amazon, and OVH; however, we also find Internet access providers (Cogent,
Comcast) with significant share. It would not be a surprise that Lightning
nodes are run on cloud or hosting servers. This has been well documented for
the Bitcoin network itself~\cite{Gencer2018}.

The data from 2020-02 seem to show a reversal of the top-ranking entries.
Cogent and Comcast together account for about 11\% of the public IPv4
addresses. However, this changes when we filter out those IP addresses that
\emph{zmap} did not identify as live. Although Cogent remains in the first
place, the remainder is made up of hosters. The tentative conclusion we offer
here is that the network is moving towards central infrastructure, while at the
same time spreading over more ASes and keeping a significant share of operators
with classic ISP-based Internet access. This finding would be very consistent
with the relatively low churn we estimate.

\textbf{Key take-aways.} The \lnn seems to grow at a moderate but steady pace.
This is a good result for overall availability as every new node will also
contribute at least one new channel. The high number of nodes that do not
propagate their network addresses is remarkable: it means that nodes that wish
to establish more channels are actually limited in their choice, which is
contrary to the goal of decentralization. The network churn is quite low,
which is a good result. We find that cloud providers feature prominently among
the most important Autonomous Systems contributing to the network, which is
indicative of centralization. The number of stale nodes is high, despite the
protocol aiming to eliminate stale nodes from the network view.  Even Tor nodes
are frequently stale, hinting at operators not desiring to make a contribution
to overall network stability and availability but rather focusing on their own
privacy in more short-lived participation. Overall, the node availability on
the network leaves considerable room for improvement as better availability
would help sustain reachability.

\section{Discussion}
\label{sec:discussion}

In the previous section, we identified a number of properties concerning the
reliability of the \lnn network in terms of availability of payment routes and
overall node availability in the network. We offer several thoughts based on
our results in the context of the overarching goal of the network, which is to
enable reliable payments while remaining decentralized. These two goals are
conflicted. In our considerations, we lean towards favouring reliable payments
over decentralization and privacy. Ultimately, this is a question of which
\textit{business model} the \lnn should operate under, which is a question for
the community to decide.

\paragraph{Ability to route payments} The ability of the network to reach nodes
via source routing (combining channels) is disappointing. Only small payments
have a reasonable chance of success: 72\% of payments of \usd{0.01} and 58\% of
\usd{1} were successful. However, these payments still reached only 35\% of
nodes at best. The success rate of payments beyond \usd{50} was very low. A key
reason for this is that channels simply do not have enough funding---transient
errors due to insufficient balances are the most common.  On the face of it,
the \lnn will need to win users who are willing to invest more funds into
channels. Naturally, this is a chicken-egg problem as lack of user base is a
disincentive for investment. Investment by companies and appropriate marketing
campaigns may be one way to address this; however, such undertakings tend to
de-emphasize decentralization.

\paragraph{Extending the protocols} A good part of our experiments failed due
to the long-lived unavailability of a node representing a channel endpoint.  As
we actually measure relatively little churn in the network, at least in the
public part of the network, the question is whether the Lightning protocol
should be improved to eliminate stale information faster. Our results show that
much stale information is kept by clients, despite a pruning mechanism in the
popular \cl client. One can conceive a probing protocol that simulates payments
(without requesting a particular amount) to test channel and node availability.
Our recommendation is to add such a maintenance protocol to the network.

\paragraph{The right threat model for privacy} The argument against revealing
public IP information is that it harms the privacy of operators, whose node IDs
can be linked to their network address. This is likely to harm availability and
actually pave the way towards more centralization as publicly known nodes are
the only ones that new, joining nodes can connect to. The unavailability of
more publicly known IP addresses also counteracts the goal of eliminating stale
information as it makes it impossible to test a node's (IP) availability
directly. We believe that the feature of hiding the network address is based on
a threat model that is inappropriate for \lnn. An IP address by itself does not
reveal the operator's identity, but a hoster or ISP will be able to map an IP
address to a real person.  Companies generally release customer data only to
law enforcement agencies or other entities that can request such information,
unless they are compromised and suffer a data breach. Hence, the only adversary
that makes sense is one who falls into one of these two categories. Also note
that, depending on the gravity of the underlying cause for a request to release
customer information, it is conceivable that the requesting entities may be
able to approach the hoster of a `hub node', \ie the node with many channels
and corresponding knowledge to which nodes and IP address they map---even if
that hub is located in another jurisdiction. In such a case, the IP addresses
of operators that have disabled the propagation of their network addresses is
also given away. In summary, we believe that hiding the network address is of
little benefit for the majority of the network while harming its development.
Ideally, payments are routed in an onion-style anwyay, which should already
achieve a high level of privacy for many users and operators.

\paragraph{Business model and centralization}

The above considerations imply that the corresponding design decisions should
depend on the business model that Lightning operators wish to follow. Although
the authors of~\cite{Beres2019} state that operating a node is currently not
economically viable, this does not rule out that a dedicated, different
business model may make Lightning both more performant and more profitable.
Better funded channels, for instance, may attract more users and operators,
hence improving the network's performance and ability to route payments, which
in turn may attract further users and investment. The community should engage
in a discussion which direction the network should take.

\section{Conclusion}

In this paper, we have contributed a study of Bitcoin's Lightning network
nearly four years after its inception. We focus on two aspects of availability
that have seen little to no previous investigation, namely availability of
payment routes and overall node availability in the network. These are
necessary ingredients for the network to achieve reliable sending of payments.
We find that we can theoretically construct paths to nearly all channels and
nodes in the network. However, in practice routing payments fails much too
often, in particular when trying to send larger payments in excess of \usd{50}.
The reasons have to do with the network's composition and slow reaction to
stale information. This leads us to consider the question which changes to the
\lnn's structure, protocols, and business model could be made to improve its
overall usefulness.  While decentralization is a worthwhile goal, some
improvements may well be easier to implement against the background of a
business model involving central hubs. However, Bitcoin's philosophy was always
one of decentralization and enabling better privacy for digital payments. This
remains a central tenet for many users and operators. We believe our findings
should at least generate a discussion which direction the network should take.

We note that research on the \lnn should not end even if a new, more
investment-based business model is adopted and leads to more centralization and
less privacy. Assuming the network begins to carry sizable volumes, it will be
an interesting research question whether attacks against hub availability
benefit the attackers sufficiently in terms of stealing channel funds or
destroying user trust, for example. As a cryptocurrency and community-based
project, a growing \lnn should be continued to be monitored.

\bibliographystyle{IEEEtran}
\bibliography{main_strip}

\end{document}